\documentstyle[10lomcon,cite]{article}

\input{psfig}
\begin{document}

\title{
ON THE INFLUENCE OF EINSTEIN--PODOLSKY--ROSEN EFFECT
ON THE DOMAIN WALL FORMATION \\
DURING THE COSMOLOGICAL PHASE TRANSITIONS
}

\author{
Yu.V.Dumin
\footnote{e-mail: dumin@cityline.ru, dumin@yahoo.com}
}

\address{
IZMIRAN, Russian Academy of Sciences, \\
Troitsk, Moscow region, 142190 Russia
}

\maketitle\abstracts{
Search for the mechanisms suppressing formation of
Higgs field defects (e.g. domain walls) during
the cosmological phase transitions is one of key
problems in explanation of the observed large-scale
uniformity of the space--time. One of the possible
solutions can be based on accounting for
Einstein--Podolsky--Rosen (EPR) correlations. Namely,
if a coherent quantum state of the Higgs field was
formed in the course of its previous evolution, then
reduction to the state with broken symmetry during
the phase transition should be correlated even at
the scales exceeding the local cosmological horizons.
Detailed consideration of the simplest
one-dimensional cosmological model with $Z_2$ Higgs field
demonstrates that, at certain parameters of
the Lagrangian, EPR-correlations really result in
a substantial suppression of spontaneous formation
of the domain walls.
}

\section{
History and Present-Day Status of the Domain Wall Problem
\protect\\
\hspace*{0em}
in Cosmology
}

The first mention about the problem of domain walls
in field theories with spontaneous symmetry breaking was done by
Bogoliubov at the conference held in Pisa (Italy) in 1964.
As was written later in his report~\cite{Bogoliubov66},
``it is hard to admit, for example, that the `phases' are the
same everywhere in the space. So it appears necessary to consider
such things as `domain structure' of the vacuum.''

The next important step was done about 10 years later by
Zel'\-dovich, Kob\-za\-rev, and Okun~\cite{Zeldovich75},
who presented a first detailed consideration of the role of
domain walls in the astrophysical context and pointed both to
the positive aspects and difficulties arising in
the cosmological models involving spontaneous symmetry breaking.

At last, the problem of domain walls and other Higgs field defects
became one of the central topics of cosmology in
the early 1980s, after appearance of the inflationary models,
whose first versions were substantially based on the dynamics of
Higgs fields in the course of their phase transitions.
One of the principal difficulties of such models was associated
with the fact that the observed region of the Universe consists of
a large number of subregions that were causally-disconnected during
the phase transition. Their stable vacuum states, in general,
should be different and, therefore, separated by the domain walls,
involving considerable energy density. But subsequent evolution and
decay of these domain walls should result in dramatic irregularity
of the space--time, contradicting to observational data.

Even after development of the ``new'' and ``chaotic'' inflation
scenarios~\cite{Linde84}, which avoid the domain wall formation
at the earliest stages of cosmological evolution,
this problem still remains important for the phase transitions
occurring at the later times and less energies.

Moreover, the problem of domain walls became even more pressing
in the late 1990s, when detailed measurements of fluctuations
in the cosmic microwave background radiation (CMBR) pointed to
the absence of contribution to the large-scale structure
from the processes of disintegration of Higgs field defects.%
\footnote{
In fact, the most realistic Higgs defects in the modern theories
of elementary particles are cosmic strings and monopoles,
associated with breaking of continuous symmetries.
Nevertheless, consideration of the domain walls (related to
a discrete symmetry breaking) is still of importance as
a simplest model.
}

There is a quite large number of works aimed at solving
the above-mentioned problem. Unfortunately, their most objectionable
feature is introduction of some arbitrary modifications
to the theory of elementary particles, having no experimental
(laboratory) confirmations.
The main aim of the present report is to describe
an alternative approach, which is based on accounting for
EPR correlations and can be justified by, at least,
some indirect laboratory experiments.

\section{
EPR Correlations and Macroscopic Quantum Phenomena:
\protect\\
\hspace*{0em}
Their Probable Role in Cosmology
}

\begin{figure}[t]
\centerline{
\psfig{%
file=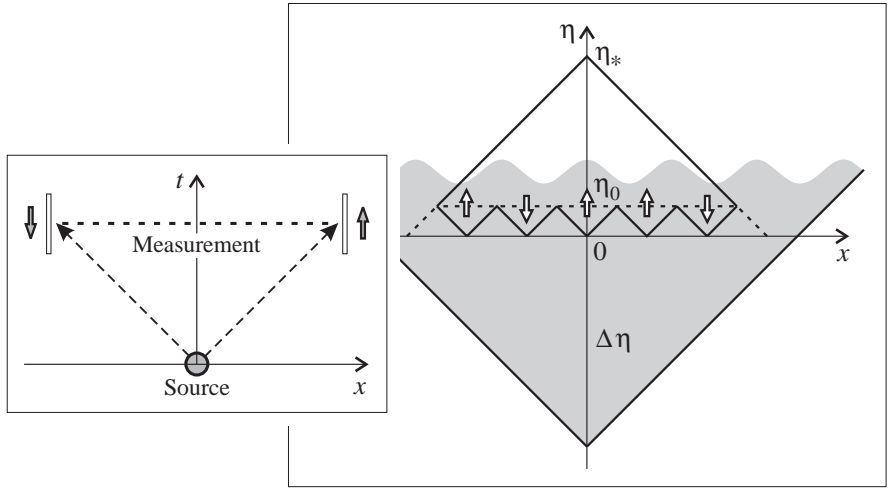%
}
}
\caption{\label{EPR-ConfDiag}
A scheme of laboratory EPR experiment (left panel) and
conformal diagram of the space--time involving a phase transition
of Higgs field (right panel).
}
\end{figure}

In the general case, Einstein--Podolsky--Rosen (EPR)
effect~\cite{Einstein35} can be defined as a correlated character
of quantum processes in the ``causally-disconnected''
(i.e., separated by a space-like interval) regions provided that
they refer to a single (coherent) quantum state of the system.%
\footnote{
It should be emphasized that EPR effect implies no violation of
the ``causality principle'': despite a correlated character,
the quantum processes in causally-disconnected regions
are random; so that no information can be transmitted
by using them.
}
The simplest and most well-known example is a correlated measurement
of the polarization states of two photons emitted by the same source,
as shown in the left panel of Fig.~\ref{EPR-ConfDiag}.

Moreover, there are some indirect experimental evidences
(the most remarkable of which were obtained in the recent years)
pointing to a possibility of EPR correlations even in
macroscopic systems. Particularly, we should mention\\
(1) the quantum-optical experiments, confirming a presence of
EPR correlations of photon pairs at considerable ($\sim$10 km)
distances;\\
(2) the experiments with ultra-cooled gases, demonstrating that
Bose condensate of a macroscopic number of particles
behaves exactly as a single coherent state,
according to all predictions of quantum mechanics; and, especially,\\
(3) the experiments on propagation of ultra-short laser pulses
through amplifying media, showing a superluminal reduction of
macroscopic coherent photon states, caused by a stimulated emission
(e.g. see review~\cite{Oraevsky98}).

On the basis of the above-listed facts, it is reasonable to assume
that EPR correlations can also manifest themselves in Bose condensate
of Higgs fields at the early stages of cosmological evolution.

\section{
One-Dimensional Cosmological Model
\protect\\
\hspace*{0em}
Involving a Phase Transition
}

To estimate efficiency of EPR effect, we shall consider
a simplest one-dimensional Friedmann--Robertson--Walker (FRW)
cosmological model with metric
\begin{equation}
ds^2 = \: dt^2 - a^2(t) \, dx^2 \,
\label{metric}
\end{equation}
and Higgs field whose Lagrangian possesses $Z_2$ symmetry group
\begin{equation}
{\cal L} \, (x, t) \, = \,
\frac{1}{2} \, \Big[ {\left( {\partial}_t \varphi \right)}^2 - \:
  {\left( {\partial}_x \varphi \right)}^2 \Big]
  \, - \,
\frac{\lambda}{4} \,
  {\Big[ {\varphi}^2 - \left( {\mu}^2 / \lambda \right) \Big] }^2 .
\label{Lagrangian}
\end{equation}

As is known, the stable vacuum states of the field~(\ref{Lagrangian}) are
\begin{equation}
{\varphi}_0 = \, \pm \, \mu \, / \sqrt{\lambda} \:\: ,
\end{equation}
and the structure of a domain wall between them is described as
\footnote{
From here on, it will be assumed that thickness of the wall
$ \, \sim 1 / \mu \, $
is small in comparison with a characteristic domain size.
}
\begin{equation}
\varphi \, (x) = \, \pm \, {\varphi}_0
  \tanh \Big[ \frac{\mu}{\sqrt{2}} \,
  ( x - x_0 ) \Big] \, ;
\label{dom_wall}
\end{equation}
so that the energy concentrated in the domain wall~(\ref{dom_wall}) equals
\begin{equation}
E = \, \frac{2 \, \sqrt{2}}{3} \, \frac{{\mu}^3}{\lambda} \; .
\label{d-w_energy}
\end{equation}

Next, by introduction of the conformal time
$ \eta = \! \int dt / a(t) \, $,
the space--time metric~(\ref{metric}) can be reduced to
the conformally flat form (e.g.~\cite{Misner69}):
\begin{equation}
ds^2 = a^2(t) \, [ \, d{\eta}^2  - dx^2 \, ] \, ;
\end{equation}
so that the light rays
($ ds^2 \! = \! 0 $)
will be described by the straight lines inclined at
$ \pm \pi / 4 \, $:
$ x = \pm \eta + {\rm const} $.

If $ \eta = 0 $  and $ \eta = {\eta}_0 $ are the beginning and end
of the phase transition, respectively,
and $ \eta = {\eta}_* $ is the instant of observation,
then, as is seen in the conformal diagram
drawn in the right panel of Fig.~\ref{EPR-ConfDiag},
\begin{equation}
N \, = \, ( {\eta}_* - {\eta}_0 ) / {\eta}_0 \,
  \approx \, {\eta}_* / {\eta}_0
\quad (\mbox{at large } N)
\label{num_subreg}
\end{equation}
is the number of spatial subregions causally-disconnected
during the phase transition. (Their final vacuum states are arbitrarily
marked by the arrows.)

A probability of phase transition without formation of
the domain walls is usually estimated as a ratio of
the number of Higgs field configurations without domain walls
to the total number of field configurations:
\begin{equation}
P_N^0 \, = \, 2 \, / \, 2^N \, ,
\end{equation}
and this quantity tends to zero very sharply at
$ N \to  \infty \, $.

On the other hand, if a sufficiently long interval of the conformal time
\begin{equation}
\Delta \eta \, \ge \, {\eta}_*
\label{coher_state}
\end{equation}
preceded the phase transition, then a coherent state of the Higgs field
(shown by the lower dashed triangle)
will be formed by the instant $ \eta = 0 $
in the entire region observable at $ {\eta}_* $
(which is shown by the upper triangle).

The inequality~(\ref{coher_state}) can be satisfied, particularly,
in the case of sufficiently long de Sitter stage
(which is typical for an overcooled state of the Higgs field
just before its phase transition). Really, if
$ \, a(t) = \exp (Ht) $,
then
\begin{equation}
\eta \, = \, - \, \frac{1}{H} \, e^{-Ht} + {\rm const} \,
  \to \, - \infty \quad \mbox{at} \quad t \, \to \, - \infty \, ;
\end{equation}
so that $ \Delta \eta $ can be quite large.

Next, if condition~(\ref{coher_state}) is satisfied,
it is reasonable to assume that EPR correlations
may occur between the all $ N $ subregions. In such case,
the probability $ P_N^0 $
should be calculated with an account of Gibbs factors
for the field configurations involving domain walls:
\begin{equation}
P_N^0 \, = \, 2 \, / \, Z \: ,
\end{equation}
where
\begin{equation}
Z \, = \: \sum_{\textstyle i=1}^N \;
  \sum_{\textstyle s_i = \pm 1}
  \exp \left\lbrace \vphantom{\frac{A}{A}} \right.
  \! - \frac{E}{T} \, \sum_{j=1}^N \,
  \frac{1}{2} \, (1 - s_{j} s_{j+1})
  \left. \vphantom{\frac{A}{A}} \right\rbrace .
\label{stat_sum}
\end{equation}
Here, $ s_j $ is the spin-like variable describing a sign of
vacuum state in $j$-th subregion,
$E$ is the domain wall energy, given by~(\ref{d-w_energy}),
and $T$ is some characteristic temperature of the phase transition.

From a formal point of view, statistical sum~(\ref{stat_sum})
is exactly the same as in the Ising model,
well studied in the condensed matter physics.
By using the respective formulas (e.g. from~\cite{Isihara71}),
a final result can be written in the form:
\footnote{
Yet another method for calculating this quantity,
based on explicit expressions for the probabilities
of field configurations with various numbers of the domain walls,
was described in our article~\cite{Dumin00}.
}
\begin{equation}
P_N^0 \, = \, \frac{2}{
  {\left[ 1 + e^{- E / T} \right]}^N + {\left[ 1 - e^{- E / T} \right]}^N
  } \: .
\label{EPR_probabil}
\end{equation}

\begin{figure}[t]
\centerline{
\psfig{%
file=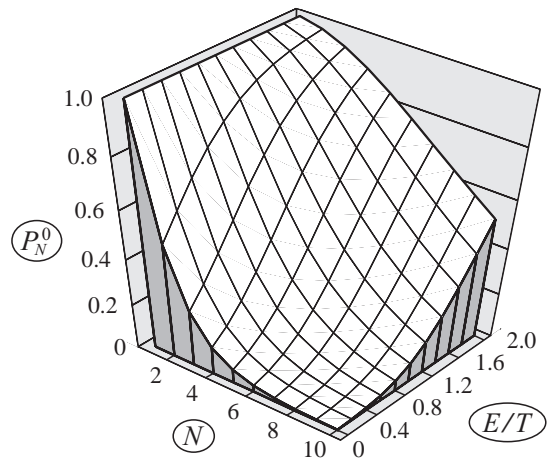%
}
}
\caption{\label{prob_distr}
A probability of phase transition without formation of
domain walls $P_N^0$ as function of the number of disconnected
subregions $N$ and the ratio of the domain wall energy to
the phase transition temperature $E/T$.
}
\end{figure}

As can be easily shown by analyzing~(\ref{EPR_probabil}),
when $ E / T $ increases,
$ P_N^0 $ becomes a very gently decreasing function of $N$.
{\it Therefore, just the large energy concentrated in the domain walls
turns out to be the factor substantially suppressing
the probability of their formation.\ }
This fact is pictorially illustrated in Fig.~\ref{prob_distr}
($ E/T = 0 $ refers to the case when
there are no EPR correlations at all).

The probability of absence of the domain walls becomes
on the order of unity (for example, 1/2) if
$ E / T \ge \ln N $,
or, with an account of (\ref{d-w_energy}) and (\ref{num_subreg}),
\begin{equation}
\frac{{\mu}^3}{\lambda \, T} \: \ge \:
  \ln \frac{{\eta}_*}{{\eta}_0} \, .
\label{d-w_absence}
\end{equation}

Because of a very weak logarithmic dependence in the right-hand side
of inequality~(\ref{d-w_absence}), this condition can be satisfied
for some particular kinds of the Lagrangians.
{\it Therefore, EPR correlations may be an efficient mechanism
of the domain wall suppression in a certain class of field theories.}

\medskip

In conclusion, it should be emphasized that the same approach,
based on accounting for EPR correlations, can be used to refine
a concentration of other Higgs field defects
(e.g., magnetic monopoles), which is one of key aspects of
the modern astroparticle physics~\cite{Klapdor97}.
The refined concentrations, in general, should be less than
the commonly accepted ones, and therefore the cosmological constraints
on the parameters of the respective field theories will be
less tight.

\section*{Acknowledgments}

I am grateful to
I.B.Khriplovich,
V.N.Lukash,
L.B.Okun,
A.I.Rez,
M.Sasaki,
A.A.Starobinsky,
A.V.Toporensky, and
G.E.Volovik
for valuable discussions, consultations, and critical comments.

\end{document}